\documentclass[%
 reprint,
 amsmath,amssymb,
 aps,
prb,
superscriptaddress,
showpacs, showkeys]{revtex4-1}

\usepackage{graphicx}
\usepackage{dcolumn}
\usepackage{bm}
\usepackage{hyperref}
\hypersetup{bookmarksnumbered, pdfpagemode=UseOutlines, 
colorlinks=true, citecolor=blue, filecolor=blue, linkcolor=blue, urlcolor=blue}

\graphicspath{/media/C6F293DFF293D253/Dropbox/1st_paper
}

\begin{document}


\title{Spin Relaxation in Graphene with self-assembled Cobalt Porphyrin Molecules}

\author{S. Omar}
\thanks{corresponding author}
\email{s.omar@rug.nl}
\affiliation{Physics of Nanodevices, Zernike Institute for Advanced Materials, University of Groningen, Nijenborgh 4, 9747 AG Groningen, The Netherlands}
\author{M. Gurram}
\affiliation{Physics of Nanodevices, Zernike Institute for Advanced Materials, University of Groningen, Nijenborgh 4, 9747 AG Groningen, The Netherlands}
\author{I.J. Vera-Marun}
\affiliation{Physics of Nanodevices, Zernike Institute for Advanced Materials, University of Groningen,  Nijenborgh 4, 9747 AG Groningen, The Netherlands}
\affiliation{School of Physics and Astronomy, The University of Manchester, Manchester M13 9PL, UK}
\author{X. Zhang}
\affiliation{Stratingh Institute for Chemistry, Zernike Institute for Advanced Materials, University of Groningen,
Nijenborgh 4, 9747 AG Groningen, The Netherlands}
\author{E.H. Huisman}
\affiliation{Physics of Nanodevices, Zernike Institute for Advanced Materials, University of Groningen, Nijenborgh 4, 9747 AG Groningen, The Netherlands}
\author{A. Kaverzin}
\affiliation{Physics of Nanodevices, Zernike Institute for Advanced Materials, University of Groningen, Nijenborgh 4, 9747 AG Groningen, The Netherlands}
\author{B.L. Feringa}
\affiliation{Stratingh Institute for Chemistry, Zernike Institute for Advanced Materials, University of Groningen,
Nijenborgh 4, 9747 AG Groningen, The Netherlands}
\author{B.J. van Wees}
\affiliation{Physics of Nanodevices, Zernike Institute for Advanced Materials, University of Groningen, Nijenborgh 4, 9747 AG Groningen, The Netherlands}%

%


\date{\today}

\begin{abstract}
In  graphene spintronics, interaction of localized magnetic moments with the electron spins paves a new way to explore the underlying spin relaxation mechanism.
A self-assembled layer of organic cobalt-porphyrin (CoPP) molecules on graphene provides a desired platform for such studies via the magnetic moments of porphyrin-bound cobalt atoms.
In this work a study of spin transport properties of graphene spin-valve devices functionalized with such CoPP molecules as a function of temperature via non-local spin-valve and Hanle spin precession measurements is reported. 
For the functionalized (molecular) devices, we observe a slight decrease in the spin relaxation time ($\tau_s$), which could be an indication of enhanced spin-flip scattering of the electron spins in graphene in the presence of the molecular magnetic moments. The effect of the molecular layer is masked  
for low quality samples (low mobility), possibly due to dominance of Elliot-Yafet (EY) type spin relaxation mechanisms. 
\begin{description}
\item[PACS numbers]
 \verb+85.75.-d+, \verb+73.22.Pr+, \verb+75.76.j+
\end{description}
\end{abstract}

\maketitle


Graphene, one atom thick layer of sp$^2$ carbon atoms, has potential for spintronic applications due to 
theoretically predicted high  spin relaxation time ($\tau_{s} \approx$ 100 ns)  and long spin diffusion length 
($\lambda_{s} \approx$ 100 $\mu$m)\cite{branas_graphene, macdonald_graphene}. These exceptional properties are attributed to
negligible spin orbit coupling and weak hyperfine interaction due to the low atomic mass of carbon \cite{dnp_magda}.
However, the  maximum reported experimental values
demonstrate  $\lambda_s$
of about 12 $\mu$m  \cite{guimaraes_controlling_2014} for encapsulated graphene and $\tau_s$ about 2.7 ns for the hydrogenated graphene
 \cite{magda_hydrogenation}, which although remarkable when compared with other metals and semiconductors, are still lower by more than an order in magnitude than the theoretically predicted values.
A mismatch  between  theory and experiments suggests towards external factors such as impurities/defects present
 near the graphene lattice, which dominate the spin relaxation process and result in a lower value for $\lambda_s$.

In order to probe the role of impurities on spin transport, one can  systematically introduce
them to graphene.  In  recent years, different research groups 
 have demonstrated  several ways of introducing impurities (magnetic and non-magnetic) in graphene
 such as  doping with adatoms, introducing defects
and chemical-functionalization  \cite{nair_point_defect,wei_chen_graphene_ferromagnetism,
butz_graphene_proton_irradiation,flipse_point_defect,zhu_fluorinated_graphene}, each method introducing a 
different spin relaxation source. For example 
heavy metal atoms such as Au can change the spin transport properties in graphene via
spin orbit coupling \cite{kawakami_Au_graphene}. On the other hand, light  metal (Mg) ions  can introduce charge 
impurity scattering of spins in graphene \cite{fabian_relaxation_sub}, although the experimental study  
rules out the role of this mechanism \cite{kawakami_Mg_graphene}. A significant change in the spin transport 
properties of graphene was reported in the presence of magnetic moments \cite{kawakami_hydrogenation}, which
can be introduced via hydrogenation or by introducing vacancies in the graphene lattice.
Remarkably, recent weak localization measurements on graphene \cite{Folk_relaxation} also show that magnetic impurities could 
play a key role in limiting the spin relaxation time in graphene. As proposed by Fabian $\textit{et al.}$ 
\cite{fabian_resonant_scattering}, if the localized moments are present at adatoms, they can act as spin hot spots and 
enhance the spin relaxation process via resonant scattering. Therefore, the recent findings serve as an imperative
to introduce magnetic impurities in graphene and investigate their 
effect on the spin transport. 

Introducing the impurities via the methods described above may damage the graphene lattice 
and modify its electronic band structure \cite{boukhvalov_graphene_destruction}. Alternatively, 
the self-assembly of molecular layers on graphene is a non-destructive way to functionalize the graphene surface 
and  one can still tune the electronic properties of this two dimensional material \cite{mallik}. Recently,
Zhang $ \textit{et al.}$  have reported the self-assembly of  porphyrin ligand bound 
cobalt atoms (CoPP) on top of a graphene surface \cite{ xiaoyan_thesis}. 
Porphyrins are attached to
graphene via weak Van der Waals interactions, while the cobalt atoms do not form any chemical bond with graphene
in contrast to the direct deposition of metal atoms or ions as discussed above \cite{hermanns}.
Therefore, the self-assembly is not supposed to change the electronic properties of graphene significantly. 
On the other hand, cobalt atoms have an unpaired spin (S=1/2), which can act as a localized magnetic moment.

In this work, we  study spin transport properties of a CoPP-graphene system
as a function of temperature, using non-local spin-valve and Hanle spin precession measurements.  After the self assembly 
of magnetic molecules, a reduced $\tau_s$ up to 50$\%$ with a lowered spin diffusion coefficient $D_s$ is obtained compared to the values for the sample without functionalization (pristine sample). A pronounced effect of the molecular layer 
was observed for samples with high mobility and high diffusion coefficient, alluding to the sample quality playing an important role in determining the spin transport properties in graphene in contrast to previous studies \cite{han_spin_2012}.

\begin{figure}
\includegraphics[scale=0.45]{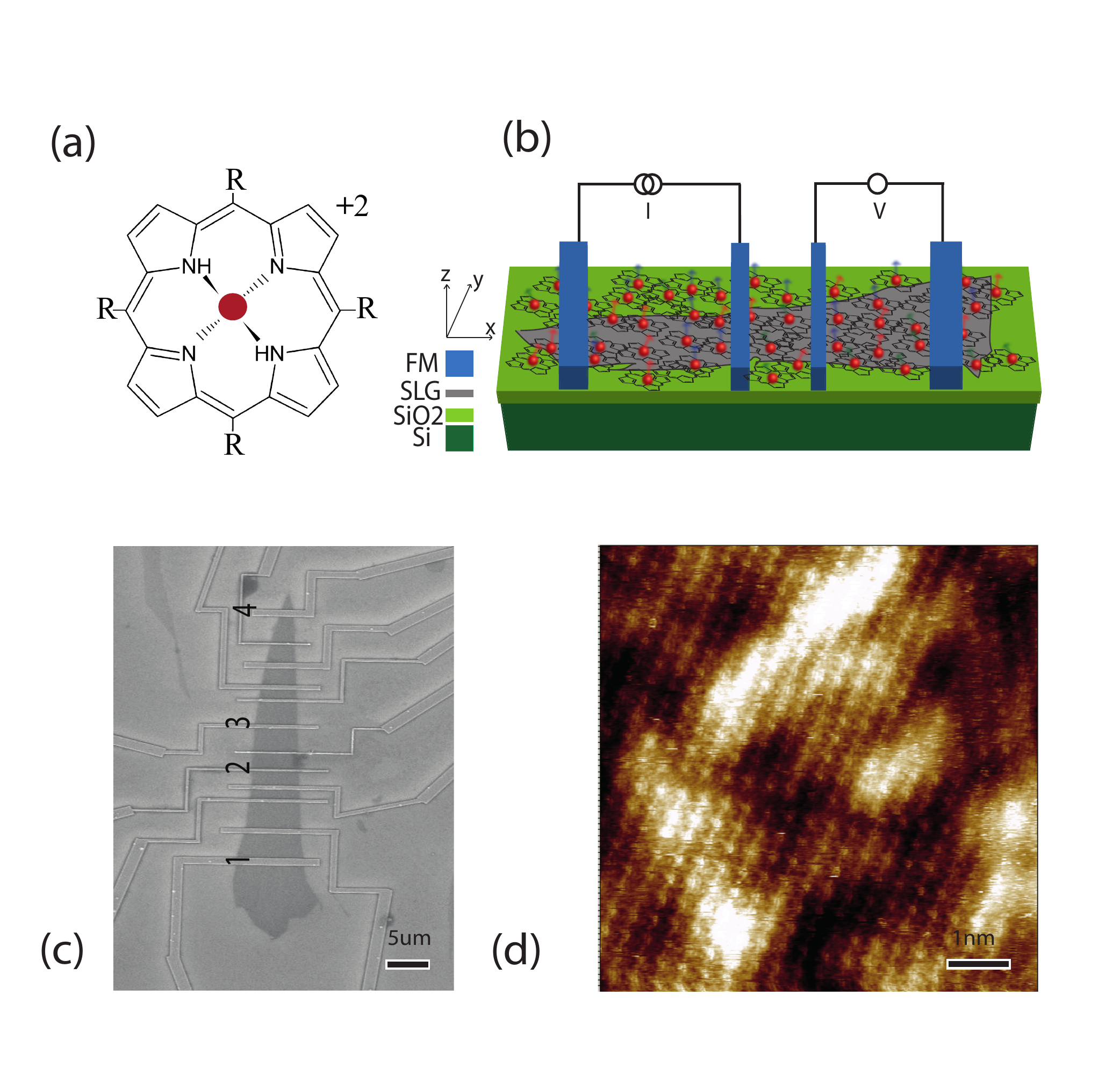}
\caption{\label{fig:fig 1}
(a)Molecular structure of a cobalt bound porphyrin (CoPP) complex. Co$^{++}$ (in red circle) is the central atom in the complex, surrounded 
by the Porphyrin ligand. In the porphyrin ring `-R' represents a long chain alkyl group (-C$_{10}$H$_{21}$), which is responsible for making 
weak Van der Waals interaction with graphene during the self-assembly.
(b) Non-local measurement scheme  for a graphene spin-valve. Graphene  (in gray) with
 a self-assembly of cobalt-porphyrin molecules on top (cobalt magnetic moments in red), is probed with ferromagnetic tunnel contacts (in blue). (c) 
 Scanning electron micro-graph (SEM) of sample A. The distance between contacts 2-3 (transport channel) is 5 $\mu$m. Outer contacts are chosen far enough from the inner ones, in order to make sure that they do not affect the spin transport.  (d) A scanning tunneling microscopy (STM) image of CVD graphene functionalized  with cobalt-porphyrin molecules on top (scan area 39 nm$^2$) on Si/SiO$_{2}$ substrate, which demonstrates an ordered self-assembly of the CoPP molecules on graphene. A bright 
 spot in the image corresponds to the core of the porphyrin molecule.}
\end{figure}

 Graphene spin-valve devices are prepared using highly oriented pyrolytic graphite (HOPG), which has a very low amount
 of impurities (ZYA grade, supplier: SPI). Graphene is mechanically 
exfoliated on to  a pre-cleaned Si/SiO$_2$ substrate (300~nm thick SiO$_2$), where 
n$^{++}$ doped Si is used as a back gate electrode. Ferromagnetic (FM) contacts 
are patterned via electron beam lithography on the PMMA coated graphene flake. Then 0.4 nm of titanium (Ti) is deposited in two steps, each step followed by oxidation to define a tunnel barrier, which
is to overcome the conductivity mismatch problem\cite{maassen_contacts}. On top of the oxide barrier we deposit 35 nm of cobalt for the spin selective
contacts. To prevent oxidation of the ferromagnetic electrodes, the contacts are covered with 3 nm thick aluminum layer followed by the lift-off process. 
A lock-in amplifier detection technique is used to measure the charge and the spin transport properties of our samples.  All the measurements are carried out using a cryostat in vacuum  ($\sim$ 1 $\times$ 10$^{-7}$ mbar) at
 different temperatures  between 4K and 300K. First, the sample is characterized in its pristine state. Afterwards, the magnetic impurities are added to the sample and the change in the charge and spin transport properties is measured. In order to equip graphene
with magnetic molecules, a cobalt-porphyrin solution (conc. 0.56 mg/ml in tetradecane) 
is drop cast on top of the device and left to dry for 10 minutes. The residual porphyrin layers on top are removed
by rinsing the device with hexane (Fig.~\ref{fig:fig 1}(b)). Since the exfoliated samples on the insulating SiO$_2$
 substrate are not big enough to perform  scanning tunneling microscopy (STM), in order to confirm the self-assembly of porphyrins on graphene we perform STM on the large area CVD graphene-CoPP system. An STM image (Fig.~\ref{fig:fig 1}(d)) of a CVD graphene sample (Si/SiO$_{2}$ substrate) with the CoPP molecules on top confirms the self-assembly of cobalt-porphyrin 
molecules on graphene. 
 
 
We report the measurements for three samples, prepared under identical conditions.
For discussion, they are labeled as sample A, B and C. A scanning electron microscope (SEM) image of sample A is shown in Figure.~\ref{fig:fig 1}(c). 
 For the charge transport measurements, an alternating current (ac) is applied between contacts 1 and 4  and the voltage is measured between contacts 2 and 3 (Fig.~\ref{fig:fig 1}(c)). In order to measure the carrier density dependence of the graphene
resistivity (Dirac curve), we sweep the back-gate
voltage.   
\begin{figure}
\includegraphics[scale=0.31,trim= 0in 0in 0in 0.0in, clip]{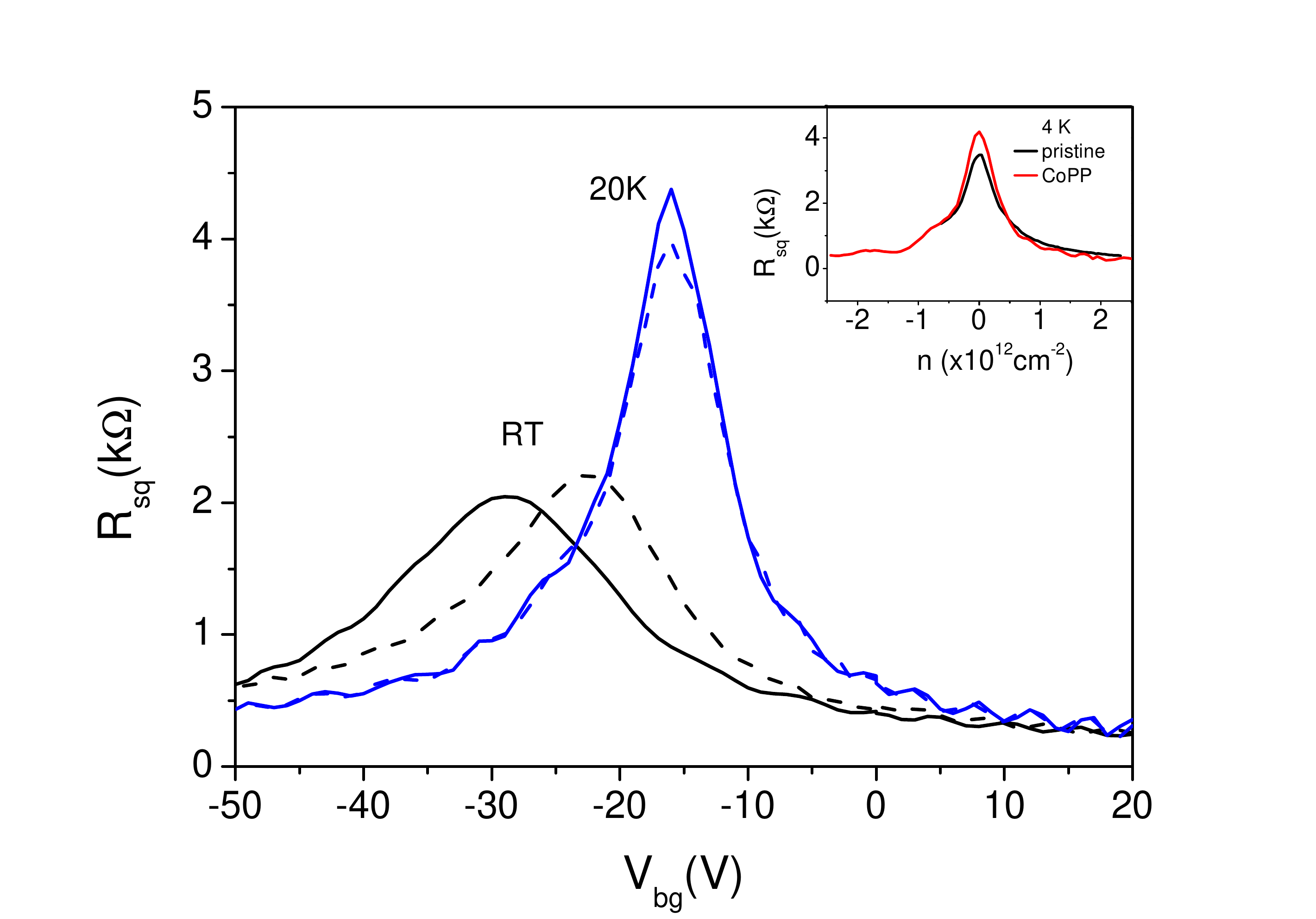}
\caption{\label{fig:fig 2}
 Resistivity as a function of gate voltage for the CoPP device (sample A) at different temperatures.  Solid (dashed) lines in the plot correspond to forward (backward) sweeping direction of the back-gate voltage. The CoPP device shows hysteresis at room temperature (black curve), which disappears at low temperatures (blue curve).  Hysteresis at RT indicates towards a
 charge transfer process between the CoPP molecules and graphene, which disappears at low temperatures due to freezing
 of the charged states in the molecules \cite{jing_legand}. A comparison between the Dirac measurement for the
 pristine and the CoPP state of the sample A is shown in the inset (at 4 K). After functionalization, the sheet resistance increases near the charge neutrality point, which is not significant at high carrier densities.}
\end{figure}
After the self-assembly of the CoPP molecules on the sample, the gate dependence is found to have a positive hysteresis at room temperature (inset Fig.~\ref{fig:fig 2}), which alludes to a charge transfer process between graphene and the CoPP molecules \cite{jing_legand,veligura_relating_2011}. At 
low temperatures charge states are frozen in the molecules and no hysteresis is observed. 
The field effect electron mobility $\mu_e$ for 
the pristine device is 7100 cm$^2$V$^{-1}$s$^{-1}$, and for the CoPP device $\mu_e \sim$ 5000 cm$^2$V$^{-1}$s$^{-1}$, both mobilities calculated at room temperature (RT) for a carrier density 
 $\sim$ 10$^{12}$ cm$^{-2}$. Contact resistances ($R_c$) for all the samples were high enough ($\geq$ 1.5 k$\Omega$) to be in the non-invasive regime as described in ref. \onlinecite{maassen_contacts}.

For the spin transport measurements, a four probe non-local detection scheme is used (Fig.~\ref{fig:fig 1}(b)).
This method allows us to decouple the charge and spin current paths and thus minimize the charge contribution
to the detected spin signal (R$_{NL}$=V$_{NL}$/I) \cite{Tombros_nature}. The spin-valve measurement is performed by
first setting a high magnetic field $\overrightarrow{B}$ 
along the -y direction (Fig.~\ref{fig:fig 1}(b)), so all the FM electrodes are magnetized along the field (parallel configuration). Then sweeping the field in the opposite direction, the electrodes reverse their magnetization at different fields  depending on their coercivity, leading to an anti-parallel
 configuration between the inner injector and the detector electrodes, which appears as a switch in the  non-local signal. 
 At high magnetic field, all the electrodes are again magnetized in the same direction in the parallel configuration. The difference between the parallel and the anti-parallel signals is the 
 spin-valve signal $\Delta R_{NL}$. 
The outer contacts are chosen  far away from the inner electrodes. In this way their influence on the measured spin signal is eliminated and we see only two distinct  switches that correspond to the magnetization reversal of the inner injector and the inner detector. 
\begin{figure}
 \includegraphics[width=\columnwidth,trim= 0in 0in 0in 0in,clip]{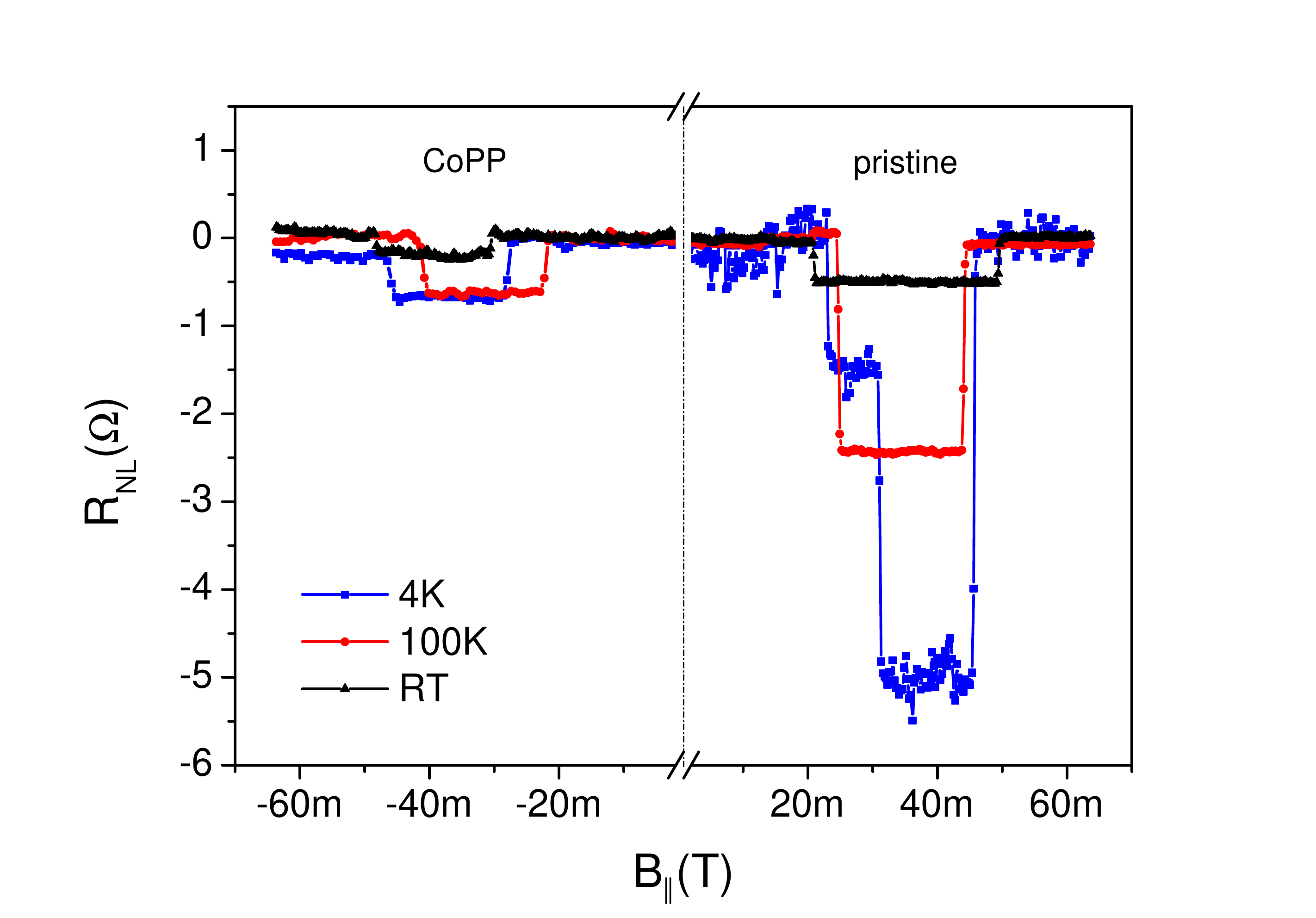}
 \caption{\label{fig:fig 3} Spin-valve measurements for sample A are shown in the positive
 x-axis for the pristine state and for the device after the functionalization are shown in the negative x-axis. A strongly reduced spin-valve signal is observed after the functionalization.}
\end{figure}

Spin-valve measurements for sample A before and after the functionalization are shown in  Fig.~\ref{fig:fig 3} at different temperatures. For both pristine and functionalized states of the sample, the spin-valve signal shows the switches corresponding to the contacts magnetization. However, after the functionalization, the signal magnitude is significantly reduced. At low temperature, the signal magnitude is increased for both
the pristine and the CoPP devices (Fig.~\ref{fig:fig 3}).

In order to understand the effect of localized magnetic moments on spin transport in graphene, we
refer to the exchange field model, explained by McCreary $\textit{et al.}$ \cite{kawakami_hydrogenation}.
In this model, an electron spin in graphene can interact with the magnetic moments via an exchange 
field $\bar{B}_{ex}$, which is the average
exchange field produced by the localized moments. $B_{ex}$ varies spatially and in time in a random fashion and  influences locally 
the Larmor precession of the diffusing spins. The effect of varying precession frequencies at different 
locations resembles the D'yakonov-Perel mechanism of spin relaxation \cite{fabian_semiconductor_spintronics} and
is responsible for an additional spin dephasing. In a spin-valve measurement, an enhanced relaxation (a reduced signal) 
is expected when the moments are randomized. As one starts applying an in-plane magnetic field, the magnetic moments try to align
themselves along the 
field and their dephasing effect gets suppressed. This feature would appear as a dip in the spin-valve signal. Within this picture, the spin relaxation rate by the fluctuating exchange field causing the dip is given by the following equation:
\begin{equation}
 \frac{1}{\tau_{ex}}=\frac{{\bigtriangleup B}^2}{\tau_c} \frac{1}{{(B_{app,y}+\bar{B}_{ex,y})}^2+{(\frac{\hbar}{g_{e}\mu_{B}\tau_c})}^2},
\label{spin relaxation}
 \end{equation}
where $\bigtriangleup {B}$ is the exchange field fluctuation magnitude, $g_e$=2 is the gyromagnetic factor of the free
 electrons, $\mu_B$ is the Bohr magneton, $\hbar$ is the reduced Plank constant 
and $\tau_c$ is the fluctuation correlation time \cite{kawakami_hydrogenation}. 
According to the formula above, the maximum relaxation (dip) in the spin-valve measurement should appear when $B_{app}$=-$\bar{B}_{ex}$. Therefore the magnetic ordering of the localized moments affects the observation of the dip. 
For paramagnetic ordering one would observe the dip around $B_{app}$=0.
On the other hand, for ferromagnetic ordering, there is a non-zero exchange field $B_{ex}$ present ($\bar{B}_{ex}\neq$0) even when no external field is applied ($B_{app}$=0). Now the dip would occur at finite external applied field and would exhibit hysteresis.

For the measured spin-valve signal for the CoPP device, we do not observe any dip, either around zero or non-zero applied field. The only clear effect of introducing  the CoPP molecules  is the reduced magnitude of the spin-valve signal. 
The observed behavior can be explained by considering the magnetic moments playing the role of spin-flip scatterers in the 
transport channel, which enhance the spin relaxation process but do not produce a measurable
effective exchange field. In order to confirm if the source of the reduced spin signal is due to an enhanced spin relaxation rate, we now need to measure the spin transport parameters via Hanle spin precession measurements.

Hanle precession measurement is a reliable tool to study the spin transport properties.
Here, a magnetic field is applied perpendicular to the direction of the 
injected spins, which precess 
around  this field $ \overrightarrow{B}$ with  Larmor frequency $ \overrightarrow{\omega_L}=g_e\mu_B  \overrightarrow{B}/\hbar$. 
While sweeping the magnetic field, due to the precession, spins can be reoriented to a direction opposite to the injected one, leading to a sign reversal in the spin signal. Simultaneously, they also  dephase and result in a lower spin accumulation at higher fields. 
The Hanle precession can be fitted with the equation \cite{Tombros_nature}:
\begin{equation}
 \bigtriangleup R_{NL} \propto \int_0^\infty \frac{1}{\sqrt{(4 \pi D_st)}} e^{\frac{-t^2}{4D_st}}cos(\omega_Lt)e^{\frac{-t}{\tau_s}}dt
,
 \label{Hanle fit}
\end{equation}
where $D_s$ is the spin diffusion coefficient, $\tau_s$ the spin relaxation time and $\omega_L$ is the Larmor frequency.

\begin{figure}[tbp]
\includegraphics*[width=\columnwidth,trim= 0in 0in 0in 0in, clip]{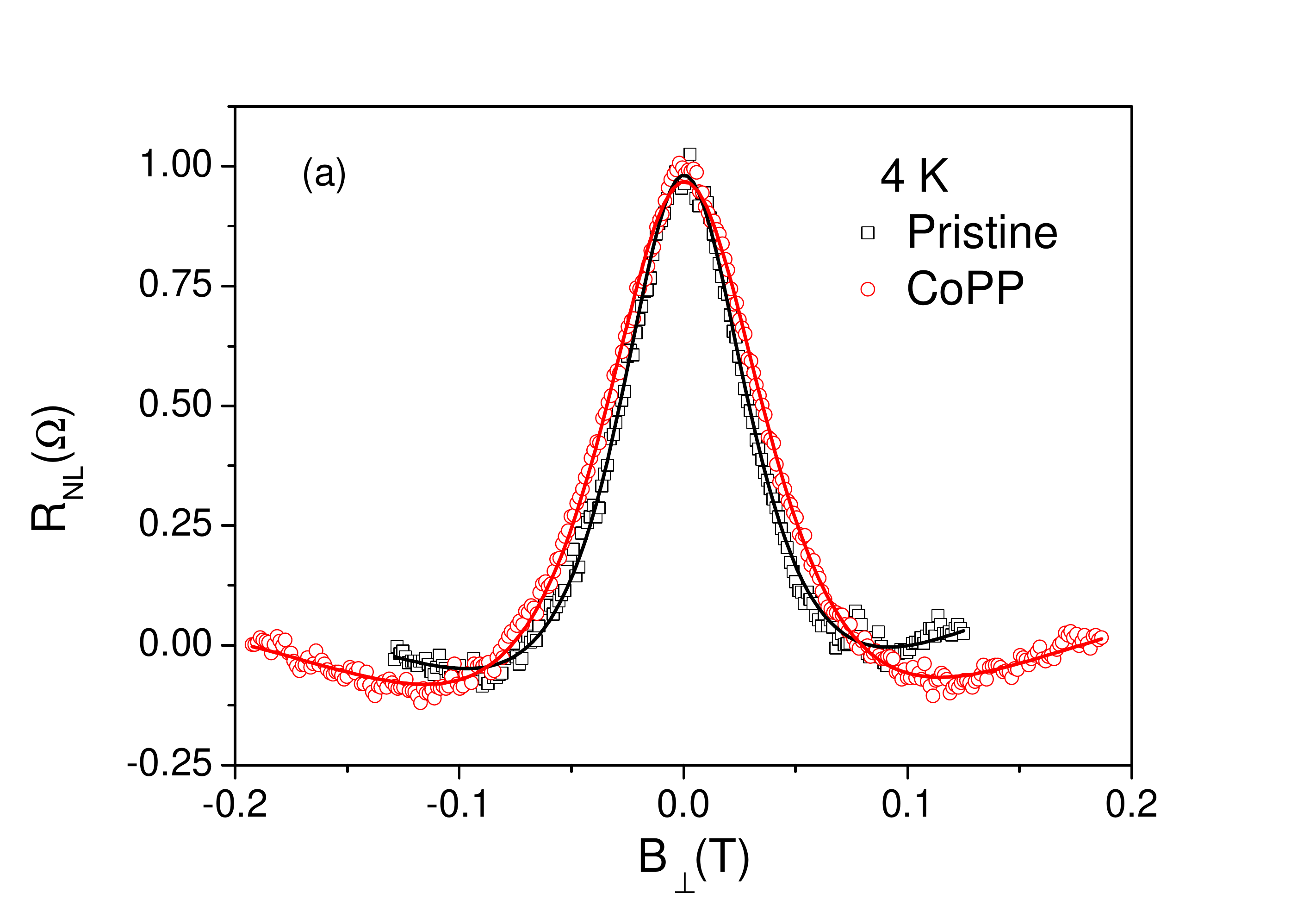}
\includegraphics*[width=\columnwidth,trim= 0in 0in 0in 0in, clip]{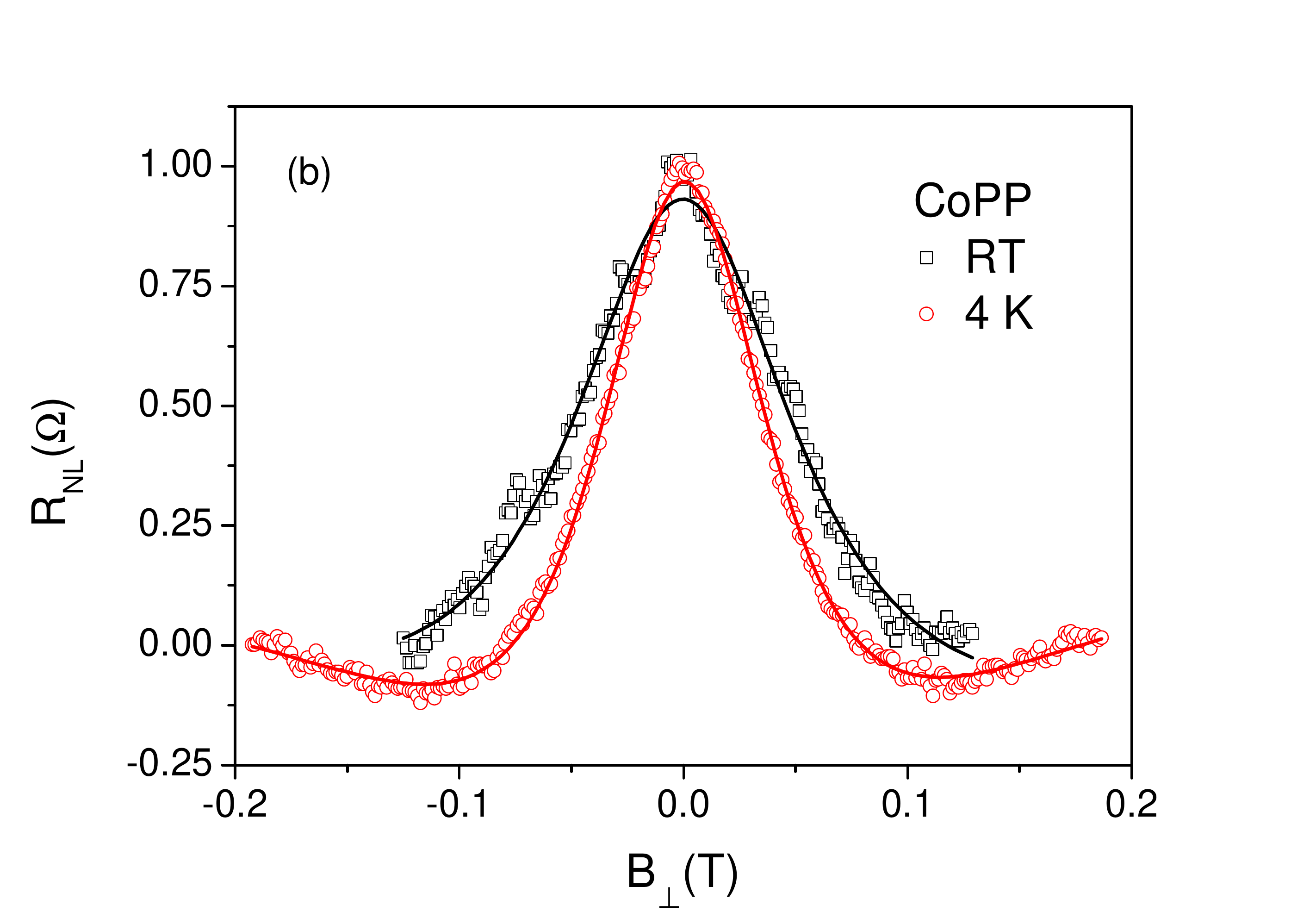}
\caption{\label{fig:fig 4}a) Hanle measurements ((R$_{P}$-R$_{AP}$)/2) for the pristine (black squares)
and the CoPP state (red circles) at 4K (sample B). The corresponding fittings are plotted in line. The curves are normalized with respect to the signal at B=0. After the functionalization Hanle line shape is broadened, indicating a reduced spin relaxation time ($\tau_s$). (b) Hanle measurements for sample B after self-assembly at RT and 4K. The curves are normalized. Broadening of the black curve (with square symbol) (RT) is dominated by the enhanced $D_s$. The spin relaxation time ($\tau_s$) only changes from 100~ps (RT) to 112~ps (4K).  }
\end{figure}

Referring back to the exchange field model, a Hanle measurement in 
the presence of an exchange magnetic field $B_{ex}$ by the magnetic moments would represent a spin precession due to a net field $B_{app}$+$B_{ex}$. The precession can result in a narrower Hanle shape due to an enhanced \textit{g} factor \cite{magda_hydrogenation, bastian_annealing,kawakami_hydrogenation} for a paramagnetic ordering of the localized moments. Whereas for the case of ferromagnetic ordering, we would expect a shifted Hanle peak.

Hanle precession curves for both pristine and CoPP devices are shown 
in Fig.~\ref{fig:fig 4}.
Here we show  the purely spin dependent signal, obtained by subtracting the anti-parallel signal from the parallel signal and the result is fitted via Eq.~\ref{Hanle fit}. The plots have been normalized to the value at $B_{app}$=0
for clearly demonstrating the change in the Hanle line shape. We observe two general trends for all measured samples. First, the Hanle curve becomes broader after 
the CoPP self-assembly. This is in contrast to the expected narrowing of the Hanle curve in the presence of a paramagnetic exchange field according to the model described above. The observed broadening indicates a reduction of the spin relaxation time, in accordance with our interpretation of the signal reduction in spin-valve measurements.  Second, upon decreasing the temperature from RT down to 4~K we do not observe any significant narrowing of the Hanle line shape which could be interpreted as an enhanced $g$ factor. On the contrary, the typical line widths and extracted spin lifetimes are not strongly dependent on temperature.

  \begin{figure}
\includegraphics[width=\columnwidth,trim= 0in 0in 0in 0in, clip]{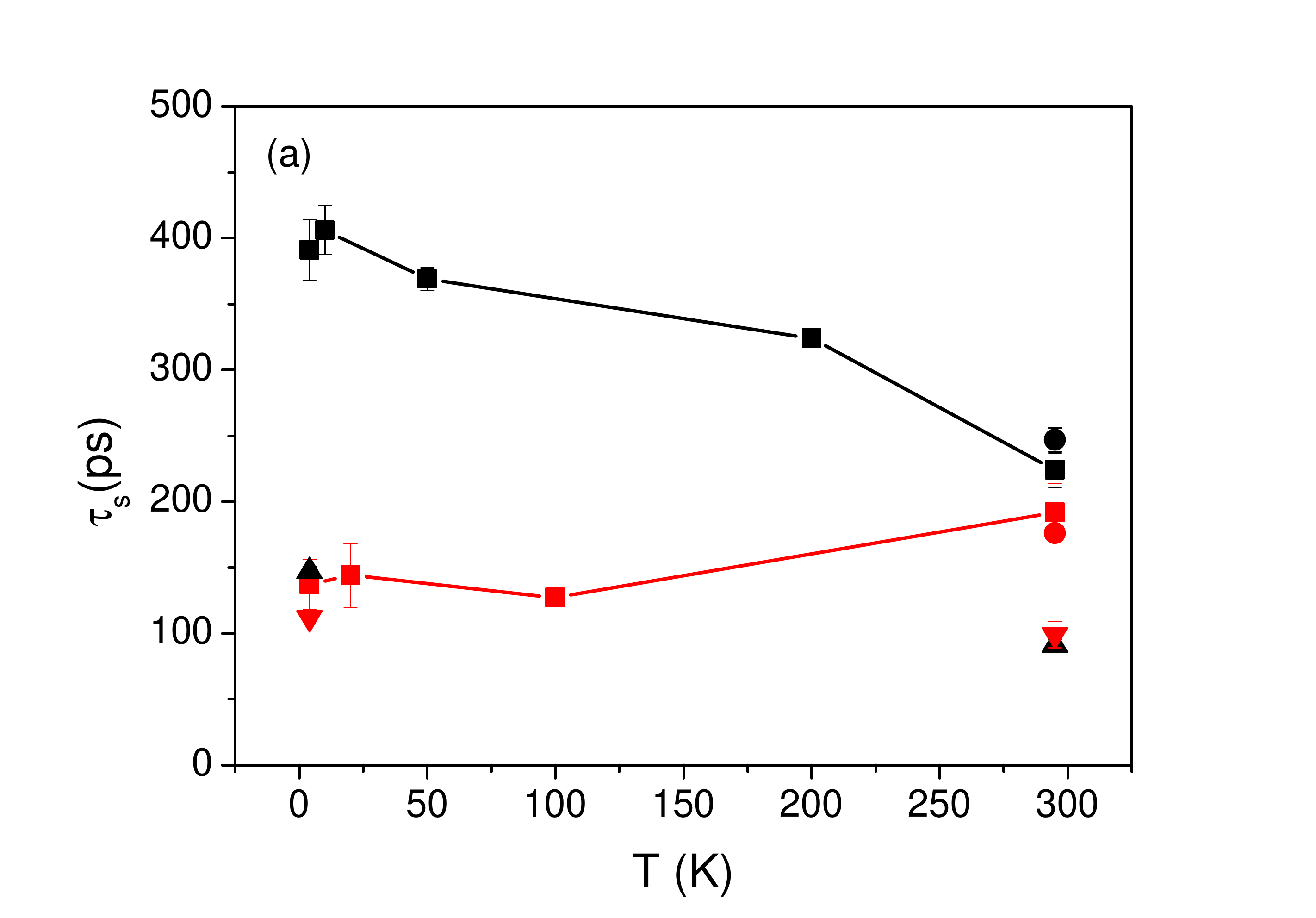}
\includegraphics[width=\columnwidth,trim= 0in 0in 0in 0in, clip]{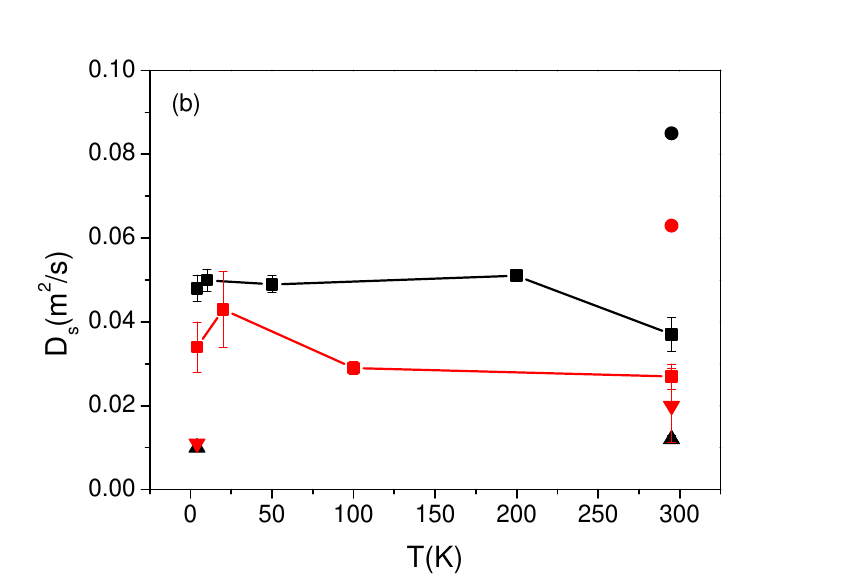}
\caption{\label{fig:fig 5}
 A summary of (a) $\tau_s$  and (b) $D_s$, extracted from Hanle analysis, for sample A (square+line), B (triangle) and C (circle) 
before (black) and after (red) the functionalization. Black data
corresponds to the pristine and red data is for the CoPP state of the samples. A reduced $\tau_s$ and $D_s$ were observed for the samples after the functionalization with a weak temperature dependence, which rules out any exchange coupling between the localized magnetic moments and the electron spins in graphene \cite{bastian_annealing} and indicates
towards an enhanced spin-flip process, where the present magnetic moments play only the role of spin-flip scatterers. The effect of the molecular layer is determined by the sample quality ($\mu_e, D$) in the pristine state. Since sample A and C have higher mobility and diffusion coefficient, $\tau_s$ is highly reduced for these samples after the functionalization. Sample B, having lower mobility did not show any significant change in $\tau_s$.  }
\end{figure}

 \begin{table}
 \vspace{0.5 cm}
\begin{ruledtabular}
\begin{tabular}{ccccc|cccc}
\multicolumn{5}{c|}{4K}&
\multicolumn{4}{c}{RT}\\
\hline
\multicolumn{1}{c}{}&pristine&
\multicolumn{1}{c}{}&CoPP&{}&
\multicolumn{1}{c}{}pristine&
\multicolumn{1}{c}{}&CoPP&{}\\
\hline
\multicolumn{1}{c}{}&{Dc}&{Ds}&
\multicolumn{1}{|c}{Dc}&{Ds}&
\multicolumn{1}{c}{Dc}&{Ds}&
\multicolumn{1}{|c}{Dc}&{Ds}\\
\hline
\multicolumn{1}{c|}{A}&{0.052}&{0.048}&
\multicolumn{1}{|c}{0.039}&{0.034}&
\multicolumn{1}{c}{0.100}&{0.037}&
\multicolumn{1}{|c}{0.050}&{0.027}\\
\hline
\multicolumn{1}{c|}{B}&{0.010}&{0.010}&
\multicolumn{1}{|c}{0.011}&{0.010}&
\multicolumn{1}{c}{0.014}&{0.012}&
\multicolumn{1}{|c}{0.020}&{0.020}\\
\hline
\multicolumn{1}{c|}{C}&{--}&{--}&
\multicolumn{1}{|c}{--}&{0.12}&
\multicolumn{1}{c}{--}&{0.085}&
\multicolumn{1}{|l}{--}&{0.063}\\
 \end{tabular}
\end{ruledtabular}
\caption{\label{tab:tab_1}A summary of $D_c$ and $D_s$ (units in $m^2/s$) for sample A, B and C, before (pristine) and after (CoPP) functionalization. For all the samples, $D_c$ and $D_s$ are approximately in the similar order. For sample A, $D_c$ in the pristine state is found around 0.05 $m^2/s$. We also sometimes observed an asymmetry in the Dirac curve at different temperatures. This asymmetry rises due to contact induced doping at different regions \cite{volmer_suppression_2014}, resulting  in a different value for $D_c$ at different temperatures.}
\end{table}


A summary of the extracted spin parameters for all samples studied in this work is presented in Fig.~\ref{fig:fig 5}. For sample A we observed the strongest effect of the molecular layer on the spin parameters. In its pristine state, the extracted spin relaxation time $\tau_s$ is in the range of 300--400~ps for all the measured temperatures, with a corresponding spin diffusion length  $\lambda_s(=\sqrt{D_s\tau_s})$ 3--4.5~$\mu$m. On the other hand, after self-assembly sample A exhibited a strongly reduced $\tau_s$ in the range 100--200~ps and a correspondingly lower $\lambda_s \sim$ 2--2.5~$\mu$m. Interestingly, we did not observe any significant temperature dependence for the extracted $\tau_s$ in the measured temperature range, which would have otherwise been expected due to the presence of an effective exchange field via localized molecular paramagnetic moments \cite{kawakami_hydrogenation, bastian_annealing}. Therefore the added  magnetic molecules seem to only increase the spin relaxation rate via the introduction of more spin-flip scattering events. 

Furthermore, we also observed a minor reduction of the extracted spin diffusion coefficient $D_s$ after self-assembly, consistent with the observed reduction in mobility as discussed in Fig.~\ref{fig:fig 2}. Note that the reliability of a Hanle fit is typically established by comparing the agreement between the extracted spin diffusion coefficient $D_s$ with the charge diffusion coefficient $D_c$ \cite{maassen_contacts, magda_hydrogenation, kawakami_hydrogenation}. The latter can be independently calculated via the resistivity of the sample at a known density of states $\nu$ using the Einstein relation $D_c = 1/R_{sq}e^2\nu$. In the absence of electron-electron interaction $D_c$ and $D_s$ should match \cite{weber_diffusion}. As shown in Table.~\ref{tab:tab_1} both parameters are in a reasonable agreement, confirming the validity of the the Hanle analysis.

It is worth mentioning that the earlier work of molecular doping on graphene \cite{han_spin_2012} did  not exhibit any measurable change in the spin transport properties of graphene, while the charge transport properties were modified. However, we find that sample quality, as determined by the magnitude of the diffusion coefficient ($D_c$, $D_s$) or electronic mobility ($\mu_e$), plays an important role on the influence that the cobalt-porphyrin molecular layer exerts on the spin transport parameters. For example, for sample B we do not observe a significant change in $D_s$ and $\tau_s$ after self-assembly. This reduced sensitivity can be attributed to its low mobility (and diffusion coefficient) which in the pristine state was $\sim$2000~cm$^2$V$^{-1}$s$^{-1}$, almost a factor of 3.5 times lower than for sample A. On the other hand, for sample C which had a comparatively better quality ($D_s \sim 0.1$~m$^2$/s) we again observed a significant reduction of 30~\% on the spin relaxation time, confirming our initial observations. 

A significant reduction in $\tau_s$, with a simultaneous moderate reduction in $D_s$, is inconsistent with the picture of localized magnetic moments creating an effective exchange field as discussed above \cite{kawakami_hydrogenation, bastian_annealing}, or a model where localized states act as spin reservoirs \cite{maassen_epitaxial}. Both models imply a significant increase of the extracted $\tau_s$ and a proportionally reduced $D_s$, which can be understood via an enhanced $g$ factor and the symmetry of the Hanle equation \cite{magda_hydrogenation, bastian_annealing}. Furthermore, both models are also expected to show a strong temperature dependence, which is not observed here.   

The reduction on the spin transport parameters indicates that the main role of the Co-porphyrin molecular layer is to act as an extra source of spin-flip scattering. This interpretation is consistent with the lack of sensitivity to the molecular layer by low-quality samples, where the initial spin relaxation rate was already large and therefore masks the relaxation process introduced by the molecular layer. In addition, the concomitant reduction in $D_s$ and $\tau_s$ observed can be partially understood by the enhanced momentum scattering introduced by the molecular layer, since in single layer graphene the leading spin relaxation mechanism is of the Elliot-Yafet type, which results in the proportionality relation $\tau_s \propto D$ \cite{maassen_linear_scaling,han_spin_2011, yang_observation_2011}.  This observation is interesting, since previous experiments rule out the role of mobility dependence of $\tau_s$ \cite{han_spin_2012} or seem to observe an opposite relation between $\tau_s$ and $\mu_e$ i.e. higher spin lifetime for lower mobility samples \cite{volmer_suppression_2014}.

To summarize, we observe a change both in the charge and spin transport properties of graphene in the presence of cobalt-porphyrin molecules. In the charge transport measurements, we observe an increase in the graphene sheet resistance after  functionalization due to their interaction with graphene via weak Van der Waals forces.  For the spin transport measurements we observe lower values of $\tau_s$ and $\lambda_s$ for the CoPP-graphene system compared to the pristine one. The measurements are not strongly temperature dependent suggesting that the additional spin relaxation can be due to the enhanced spin scattering from the magnetic moments, accompanied by the other relaxation sources.  The changes are also sensitive to the sample quality ($D,\mu_e$) in the pristine state and are masked for a lower value of the mobility or diffusion coefficient, indicating also the presence of Elliot-Yafet type spin relaxation mechanism. 

We acknowledge  J. G. Holstein, H. M. de Roosz and H. Adema for their technical assistance. We would like to thank  M.H.D. Guim\~{a}raes and J.J. van den Berg for their help in sample preparation. This research work was financed under EU-graphene flagship program (637088) and supported by the Zernike Institute for Advanced Materials, the Ministry of Education Culture and Science (Gravitation program 024.601035,B.L.F.) and the Netherlands Organization for Scientific Research (NWO).
  


\end{document}